\documentclass[fleqn,usenatbib,twocolumn]{mnras}

\usepackage{newtxtext,newtxmath}
\usepackage{orcidlink}
\usepackage[T1]{fontenc}

\DeclareRobustCommand{\VAN}[3]{#2}
\let\VANthebibliography\thebibliography
\def\thebibliography{\DeclareRobustCommand{\VAN}[3]{##3}\VANthebibliography}

\usepackage{placeins}
\usepackage{graphicx}	
\usepackage{amsmath}	

\title[Nuclear Cosmochronology]{Impact of Decay-Chain Feeding by Long-Lived Heavy Nuclei on Nuclear Cosmochronology}

\author[Chen et al.]{
	Qiu-Hong Chen \orcidlink{0009-0006-8625-5283} $^{1}$,
	Meng-Hua Chen \orcidlink{0000-0001-8406-8683} $^{1}$\thanks{E-mail: mhc@gxu.edu.cn (MHC)},
	Xianlin Yang \orcidlink{0000-0002-4917-4808} $^{2}$,
	and En-Wei Liang \orcidlink{0000-0002-7044-733X} $^{1}$
	\\
	$^{1}$Guangxi Key Laboratory for Relativistic Astrophysics, School of Physical Science and Technology, Guangxi University, Nanning 530004, China\\
	$^{2}$Laboratory of Karst Environmental Evolution and Ecological Security, Institute of Geochemistry, Chinese Academy of Sciences, Guiyang 550081, China
}

\date{Accepted XXX. Received YYY; in original form ZZZ}
\pubyear{\the\year{}}

\begin{document}
	\label{firstpage}
	\pagerange{\pageref{firstpage}--\pageref{lastpage}}
	\maketitle
	
	\begin{abstract}
		Nuclear cosmochronology provides an independent method for determining the ages of astrophysical objects without relying on Galactic chemical evolution models. Long-lived radioactive actinides synthesized via the rapid neutron-capture process ($r$-process) particularly thorium and uranium serve as ideal chronometric nuclei. However, predicting the initial abundances of these nuclei and their subsequent long-term evolution remains a major source of uncertainty in nuclear cosmochronology. In this work, we investigate the impact of decay-chain feeding from long-lived progenitor nuclei on nuclear cosmochronology and estimate the corresponding age corrections. We develop a self-consistent radioactive decay network that includes all relevant decay pathways and couple it to $r$-process nucleosynthesis calculations. By tracking the long-term evolution of $r$-process nucleosynthesis based on neutron star merger trajectories, the dominant long-lived nuclei shaping the evolution of the chronometric elements Th and U are identified as $^{232}$Th, $^{235}$U, $^{236}$U, $^{238}$U, and $^{244}$Pu. It is found that delayed feeding substantially modifies the long-term evolution of actinides particularly for $^{232}$Th, which receives significant late-time contributions from heavier progenitor nuclei such as $^{244}$Pu and $^{236}$U. As a consequence, the U/Th and Th/X chronometers exhibit significant sensitivity to delayed feeding from the progenitor nuclei $^{244}$Pu and $^{236}$U, with age corrections reaching $\sim5$~Gyr. These results suggest that decay-chain feeding from long-lived progenitor nuclei plays a crucial role in nuclear cosmochronology and should be properly accounted for in age determinations based on actinide chronometers.
	\end{abstract}
	
	
	
	\section{Introduction}
	\label{sec:introduction}
	
	The astrophysical origin of the heavy elements remains a major problem in nuclear astrophysics \citep{Cowan2021RvMP}. Rapid neutron capture which is known as the $r$-process is believed to be responsible for the synthesis of heavy nuclei beyond iron \citep{Burbidge1957RvMP}. This process requires environments with high neutron densities and suitable thermodynamic conditions. Neutron star mergers are now regarded as one of the most promising sites because they eject extremely neutron-rich matter \citep{Shibata2019}. Observations of kilonova emission following the gravitational-wave event GW170817 have provided strong evidence that heavy elements are synthesized in such environments \citep{Kasen2017Nature}.
	
	Long-lived radioactive nuclei synthesized in the $r$-process provide a powerful tool for age determination \citep{Fowler1960,Schramm1970ApJ,Wene1975A&A,Thielemann1983A&A,Meyer1986, Cowan1987ApJ,Cowan1991PhR}. Nuclear cosmochronology compares observed abundances with predicted initial production ratios to infer the time elapsed since nucleosynthesis. The U/Th chronometer is of particular importance because both nuclei are produced predominantly in the $r$-process and their half-lives are comparable to Galactic and cosmological timescales. This method has been widely applied to metal-poor stars that preserve the chemical signatures of early nucleosynthesis events \citep{Cowan1999ApJ,MP:Hill2002,MP:Cowan2002,MP:Frebel2007,MP:Hill2017,MP:Placco2017,MP:Holmbeck2018,MP:Roederer2024}. It provides an independent method for age determination without relying on Galactic chemical evolution models.
	
	A key uncertainty in nuclear cosmochronology arises from the determination of the initial production ratios. This uncertainty primarily stems from the fact that the $r$-process involves a vast number of neutron-rich nuclei whose properties remain largely unmeasured, thereby requiring nuclear physics models as essential inputs \citep{Kajino2019, Chen2025A&A}. Previous studies have made significant efforts to estimate production ratios using $r$-process nucleosynthesis calculations with different nuclear mass models, nuclear reaction rates, decay properties, and astrophysical conditions \citep{Niu2009PhRvC,Niu2013PhLB,Niu2019PhRvC,Wu2022ApJ}. \cite{Wu2022ApJ} proposed a Th-U-X chronometer that constrains $r$-process conditions by combining Th/X, U/X, and U/Th ratios, thereby reducing the astrophysical uncertainties in age estimates. However, the delayed contribution from long-lived progenitor nuclei through decay chains has not been extensively explored in nuclear cosmochronology. These long-lived nuclei can undergo successive decays through long decay chains and can continue to feed the actinide region over long timescales. In addition, previous studies often considered core-collapse supernovae as the primary site of $r$-process nucleosynthesis, while more recent observational evidence suggests that neutron star mergers are a more promising source of actinides such as thorium and uranium \citep{Chen2024MNRAS}. Therefore, it is important to investigate neutron star mergers as a source of actinides and to quantify the impact of decay-chain feeding by long-lived progenitor nuclei on nuclear cosmochronology and its associated uncertainties.
	
	In the pioneering work of \cite{Fowler1960}, the initial production ratios were predicted by simply counting the number (i.e., species) of progenitor nuclei. \cite{Cowan1987ApJ} employed $r$-process nucleosynthesis calculations based on the limited nuclear data available at the time and derived the production ratios by summing progenitor abundances along $\alpha$-decay chains. However, treatments that consider only $\alpha$-decay chains are incomplete, as progenitor nuclei can also feed chronometric nuclei through decay pathways involving both $\alpha$- and $\beta$-decays.
	In this work, we investigate the impact of decay-chain feeding by long-lived heavy nuclei on nuclear cosmochronology. We perform $r$-process nucleosynthesis calculations that include a detailed treatment of radioactive decay chains. We track the long-term evolution of heavy element abundances over cosmological timescales and evaluate how production ratios are modified by delayed feeding. Our results show that the decay chains of long-lived heavy nuclei play a critical role in chronometric age estimates.
	
	This paper is organized as follows. In Section~\ref{sec:method}, we present the theoretical framework of nuclear cosmochronology and describe how the contribution of decay-chain feeding is incorporated into the calculations. Section~\ref{sec:result} presents the abundance analysis results obtained with the inclusion of decay-chain feeding and discusses its impact on nuclear cosmochronology. In Section~\ref{sec:conclusion} we summarize our main conclusions.

	\section{Methods}
	\label{sec:method}

	In order to investigate the impact of decay-chain feeding from long-lived heavy nuclei on nuclear cosmochronology, we construct a radioactive decay network that self-consistently tracks the contributions of all progenitor nuclei to chronometric isotopes.
	In Section~\ref{subsec:cosmochronology}, we outline the theoretical framework of nuclear cosmochronology, with a focus on U/Th chronometric relations. We then develop an extended chronometer framework that incorporates decay-chain feeding from all progenitor nuclei (Section~\ref{sec:radioactive}). Finally, we perform $r$-process nucleosynthesis calculations to derive the initial production ratios that serve as inputs to nuclear cosmochronology (Section~\ref{sec:nucleosynthesis}).

	\subsection{Nuclear Cosmochronology}
	\label{subsec:cosmochronology}
	
	The theoretical framework of nuclear cosmochronology is based on the exponential law of radioactive decay,
	\begin{equation}
		Y(t)=Y_0 e^{-\lambda t},
	\end{equation}
	where $Y(t)$ is the abundance of a radioactive nuclide at time $t$, $Y_0$ is its initial abundance, and $\lambda=\ln 2/t_{1/2}$ is the decay constant, with $t_{1/2}$ being the half-life. In practice, it is more robust to consider abundance ratios rather than absolute abundances.
	For a radioactive nuclide relative to a stable or long-lived reference nuclide, the abundance ratio evolves as
	\begin{equation}
		\frac{Y_A (t)}{Y_B (t)} = \frac{Y_{A,0}}{Y_{B,0}} e^{-(\lambda_A-\lambda_B)t},
	\end{equation}
	where $A$ and $B$ denote the radioactive nuclide and the reference nuclide, respectively.
	Thus, the age of astrophysical objects can be estimated by comparing the observed abundance ratios with their initial production ratios,
	\begin{equation}
		t = \frac{\ln10}{\lambda_A-\lambda_B} \left[ \log_{10}\left(\frac{Y_A}{Y_B} \right)_{\rm ini} - \log_{10}\left(\frac{Y_A}{Y_B} \right)_{\rm obs} \right],
	\end{equation}
	where the subscripts "ini" and "obs" denote the initial production ratio at the end of the nucleosynthesis event and the currently observed abundance ratio, respectively.
	
	The most commonly used chronometer pairs are U/Th, U/X, and Th/X, where X denotes a stable or effectively stable $r$-process element, typically heavier than Ba such as Eu. In the classical treatment, each chronometric nucleus is assumed to evolve as an isolated decay process, without accounting for any later replenishment by radioactive progenitors. Under this approximation, the age can be written as
	\begin{equation}
		\begin{aligned}
			t_{\mathrm{U/Th}} &\approx
			21.80~\mathrm{Gyr}
			\left[
			\log_{10}\left(\frac{Y_{\mathrm{U}}}{Y_{\mathrm{Th}}}\right)_{\mathrm{ini}}
			-
			\log_{10}\left(\frac{Y_{\mathrm{U}}}{Y_{\mathrm{Th}}}\right)_{\mathrm{obs}}
			\right], \\
			t_{\mathrm{U/X}} &\approx
			14.84~\mathrm{Gyr}
			\left[
			\log_{10}\left(\frac{Y_{\mathrm{U}}}{Y_{\mathrm{X}}}\right)_{\mathrm{ini}}
			-
			\log_{10}\left(\frac{Y_{\mathrm{U}}}{Y_{\mathrm{X}}}\right)_{\mathrm{obs}}
			\right], \\
			t_{\mathrm{Th/X}} &\approx
			46.67~\mathrm{Gyr}
			\left[
			\log_{10}\left(\frac{Y_{\mathrm{Th}}}{Y_{\mathrm{X}}}\right)_{\mathrm{ini}}
			-
			\log_{10}\left(\frac{Y_{\mathrm{Th}}}{Y_{\mathrm{X}}}\right)_{\mathrm{obs}}
			\right].
		\end{aligned}
		\label{eq:classical_chronometers}
	\end{equation}

	In fact, the reliability of these nuclear chronometers depends critically on two assumptions. The first is that the initial production ratio can be predicted reliably from $r$-process nucleosynthesis calculations. The second is that the subsequent abundance evolution of the chronometric nuclei is described solely by their own decay constants. While the first issue has been extensively studied in the literature, the second assumption is an idealization. In reality, the abundances of Th, U, and possibly also some reference elements can be modified after freeze-out by delayed feeding from heavier radioactive progenitors.
	This effect is particularly relevant for actinides synthesized in extremely neutron-rich ejecta. Heavy nuclei produced beyond Th and U may decay back into the actinide region through a combination of $\alpha$-decays and $\beta$-decays. As a result, the abundance of a chronometric nucleus at late times is not determined solely by its own direct survival, but also by cumulative feed-down from the entire decay network. Therefore, when long-lived progenitor nuclei make a significant contribution, the classical formulae above can provide only an approximate description.

	\subsection{Decay-Chain Feeding in Nuclear Cosmochronology}
	\label{sec:radioactive}

	\begin{figure}
		\centering
		\includegraphics[width=1.0\linewidth]{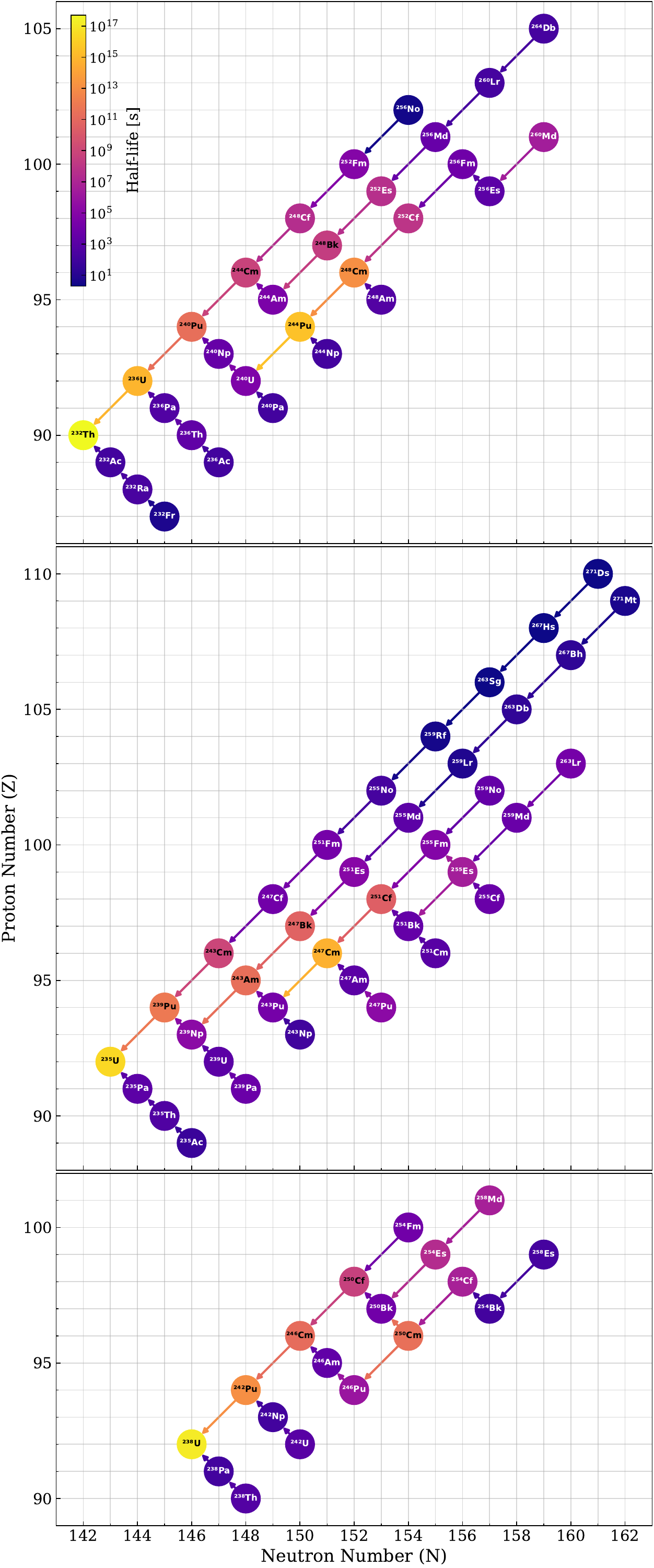}
		\caption{Dedecay networks of progenitor nuclei feeding the actinide chronometers $^{232}$Th (top), $^{235}$U (middle), and $^{238}$U (bottom), together with the decay networks feeding these nuclei. Symbols represent individual isotopes, arrows indicate the direction of radioactive decay, and the colour scale denotes the nuclear half-life in seconds.}
		\label{fig:chains}
	\end{figure}
	
	The classical formulation of nuclear cosmochronology treats each chronometric nuclide as an isolated radioactive species, with its abundance governed solely by its intrinsic decay constant. However, this approximation neglects a crucial physical process: heavy nuclei synthesized in the $r$-process undergo extended decay cascades that continuously replenish the abundances of long-lived actinides over cosmological timescales.
	In particular, nuclei with mass numbers $A \gtrsim 230$ do not decay directly into the chronometric isotopes. Instead, they evolve through sequences of $\alpha$- and $\beta$-decays involving multiple intermediate nuclei. Figure~\ref{fig:chains} shows the radioactive decay chains relevant to the Th and U chronometers. These decay chains form a complex network that redistributes material across the actinide region. Consequently, the late-time abundance of a given chronometric nucleus depends not only on its direct survival but also on cumulative contributions from all progenitor nuclei connected through the decay network. This interdependence invalidates the implicit assumption of isolated exponential decay and motivates a more comprehensive theoretical description.
	
	To quantify the decay-chain feeding effect in a self-consistent manner, we model the full radioactive decay network following the termination of nucleosynthesis. Let $Y_i(t)$ denote the abundance of nuclide $i$ at time $t$. The nuclides are ordered according to the decay flow, such that parent nuclei have smaller indices than their daughter nuclei. Including all possible feeding channels from upstream progenitors, the time evolution of the abundance distribution is written as
	\begin{equation}
		\begin{cases}
			\frac{\mathrm{d}Y_{1}(t)}{\mathrm{d}t} &= -\lambda_{1}Y_{1}(t),                              \\ 
			\frac{\mathrm{d}Y_{2}(t)}{\mathrm{d}t} &= b_{1 \rightarrow 2}\lambda_{1}Y_{1}(t) -\lambda_{2}Y_{2}(t),         \\
			\frac{\mathrm{d}Y_{3}(t)}{\mathrm{d}t} &= b_{1 \rightarrow 3}\lambda_{1}Y_{1}(t) + b_{2 \rightarrow 3}\lambda_{2}Y_{2}(t) -\lambda_{3}Y_{3}(t),         \\
			&\vdots  \\
			\frac{\mathrm{d}Y_{i}(t)}{\mathrm{d}t} &= b_{1 \rightarrow i}\lambda_{1}Y_{1}(t) +  \cdots + b_{i-1 \rightarrow i}\lambda_{i-1}Y_{i-1}(t) -\lambda_{i}Y_{i}(t),         \\
			&\vdots                                           \\
			\frac{\mathrm{d}Y_{n}(t)}{\mathrm{d}t} &= b_{1 \rightarrow n}\lambda_{1}Y_{1}(t) + \cdots + b_{n-1 \rightarrow n}\lambda_{n-1}Y_{n-1}(t)  -\lambda_{n}Y_{n}(t).
		\end{cases}
	\end{equation}
	Here, $\lambda_i$ is the total decay constant of nuclide $i$, and $b_{j \rightarrow i}$ is the branching ratio for the decay channel from nuclide $j$ to nuclide $i$. 
	The radioactive decay data are taken from the Evaluated Nuclear Data File library ENDF/B-VIII.1 \footnote{\url{https://www-nds.iaea.org/public/download-endf/ENDF-B-VIII.1/decay/}} \citep{Nobre2025ENDF}.
	In this formulation, the final term on the right-hand side describes the depletion of species $i$ through its own radioactive decay, whereas the remaining terms collectively represent the feeding contributions from all progenitor nuclei. We note that this equation can be regarded as a general formulation, since most heavy nuclei typically have only one or two dominant decay branches.
	
	For realistic decay networks involving a large number of unstable nuclei, it is advantageous to rewrite the coupled system in matrix form, which provides a compact framework for both numerical integration and formal analytical treatments.
	The decay equations can therefore be expressed as
	\begin{equation}
		\frac{\mathrm{d}}{\mathrm{d}t}
		\begin{bmatrix}
			Y_{1}(t) \\
			Y_{2}(t) \\
			Y_{3}(t) \\
			\vdots   \\
			Y_{n}(t) 
		\end{bmatrix} =
		\begin{bmatrix}
			-\lambda_{1}  \\
			b_{1 \rightarrow 2}\lambda_{1} & -\lambda_{2}  \\
			b_{1 \rightarrow 3}\lambda_{1} & b_{2 \rightarrow 3}\lambda_{2} & -\lambda_{3}  \\
			\vdots            & \vdots            & \vdots            & \ddots   \\
			b_{1 \rightarrow n}\lambda_{1} & b_{2 \rightarrow n}\lambda_{2} & b_{3 \rightarrow n}\lambda_{3} & \cdots   & -\lambda_{n} 
		\end{bmatrix} \times
		\begin{bmatrix}
			Y_{1}(t) \\
			Y_{2}(t) \\
			Y_{3}(t) \\
			\vdots   \\
			Y_{n}(t)
		\end{bmatrix}
	\end{equation}
	
	Introducing the abundance vector $\boldsymbol{Y}(t)$ and the decay transition matrix $\boldsymbol{\Lambda}$, the decay-chain feeding equations can be further written in the compact form
	\begin{equation}
		\frac{\mathrm{d}\boldsymbol{Y}(t)}{\mathrm{d}t}
		=
		\boldsymbol{\Lambda}\,\boldsymbol{Y}(t),
		\label{eq:matrix_decay_equation}
	\end{equation}
	where $\boldsymbol{Y}(t)$ is the abundance vector and $\boldsymbol{\Lambda}$ is the decay transition matrix with elements
	\begin{equation}
		\Lambda_{ij} =
		\begin{cases}
			-\lambda_i, & i=j,\\
			b_{j \rightarrow i}\lambda_j, & i > j,\\
			0, &  {\rm otherwise}.
		\end{cases}
	\end{equation}
	
	This matrix embodies the structure of a linear network operator: the diagonal elements encode the total removal rates of individual nuclides, while the off-diagonal elements capture all decay-induced transitions between distinct species. Within this representation, each column of $\boldsymbol{\Lambda}$ specifies how the decay of a given parent nucleus redistributes its abundance among all possible daughter nuclei, thereby ensuring a consistent representation of the full decay topology.
	
	The formal solution to Eq.~(\ref{eq:matrix_decay_equation}) is given by the matrix exponential,
	\begin{equation}
		\boldsymbol{Y}(t)
		=
		e^{\boldsymbol{\Lambda}t}\,\boldsymbol{Y}_{\mathrm{ini}},
		\label{eq:matrix_solution}
	\end{equation}
	where $\boldsymbol{Y}_{\mathrm{ini}}$ is the initial abundance vector at the end of the $r$-process. The propagator $e^{\boldsymbol{\Lambda}t}$ effectively resums all possible decay pathways, including multi-step and multi-branch chains, and thus provides a complete description of both direct decay and delayed feeding.
	
	Based on the time dependent abundance vector, we determine the stellar age corresponding to an observed abundance ratio by solving the nonlinear equation
	\begin{equation}
		F(t)
		\equiv
		\log \epsilon^{\rm th}_{A/B}(t)
		-
		\log \epsilon^{\rm obs}_{A/B}
		=0,
		\label{eq:root_equation}
	\end{equation}
	where $\log \epsilon_{A/B} = \log_{10}(Y_A/Y_B)$ is the logarithmic abundance ratio between elements $A$ and $B$. The superscripts ``th'' and ``obs'' denote the theoretical and observed abundance ratios, respectively.
	For the chronometer pairs considered in this study, the function $F(t)$ exhibits monotonic behavior over astrophysically relevant timescales. This property reflects the dominance of net radioactive decay, with decay-chain feeding introducing a smooth but subdominant modulation. Consequently, we determine the root of Eq.~(\ref{eq:root_equation}) robustly using standard numerical algorithms.
	
	In contrast to the classical formulation, our framework explicitly accounts for the time-dependent redistribution of abundances along decay chains. This inclusion leads to systematic deviations from simple exponential evolution and introduces quantifiable corrections to chronometric age estimates.
	To quantify these effects, we define the age correction as
	\begin{equation}
		\Delta t_{\mathrm{age}} = t_{\mathrm{DC}} - t_{\mathrm{CL}},
	\end{equation}
	where $t_{\mathrm{CL}}$ is evaluated using the classical two-isotope approximation, while $t_{\mathrm{DC}}$ is obtained by including decay-chain feeding. The initial production ratio is normalized such that $t_{\mathrm{CL}} = 13.8$~Gyr, consistent with the age of the Universe.
	The quantity $\Delta t_{\mathrm{age}}$ therefore represents the age correction relative to the classical approximation, or equivalently the systematic offset caused by neglecting decay-chain feeding.

	\subsection{Nucleosynthesis}
	\label{sec:nucleosynthesis}

	To obtain the initial production ratios essential for nuclear cosmochronology, we perform $r$-process nucleosynthesis calculations using the nuclear reaction network code \texttt{WinNet} \footnote{\url{https://nuc-astro.github.io/WinNet/index.html}} \citep{Reichert2023ApJS}. Our network encompasses 7583 nuclei, extending up to $^{337}$Og, and incorporates all relevant charged-particle reactions, neutron captures, photodisintegrations, weak interactions, and fission processes necessary for a self-consistent description of heavy element synthesis.
	Experimental nuclear masses and decay properties are adopted from AME2020 \citep{Wang2021CPC} and NUBASE2020 \citep{Kondev2021CPC} wherever such data are available. For nuclei lacking experimental information, theoretical properties based primarily on the FRDM2012 mass model are employed \citep{Moller2016ADNDT}. Neutron capture rates are calculated with the Hauser--Feshbach code \texttt{TALYS} \citep{Koning2023EPJA}, while the inverse photodisintegration rates are derived through detailed balance. Theoretical $\beta$-decay rates are taken from \citet{Moller2019ADNDT}. Fission rates are adopted from \citet{Panov2005NuPhA}, \citet{Panov2010A&A}, and \citet{Khuyagbaatar2020NuPhA}, with fission barriers taken from \citet{Moller2015PRC} and fragment distributions prescribed following \citet{Kodama1975NPA}. The remaining thermonuclear reaction rates are taken from the REACLIB compilation \citep{Cyburt2010ApJS}. For temperatures exceeding $0.01$~GK, theoretical weak rates with temperature and density dependence are included following \citet{Fuller1985ApJ}, \citet{Oda1994ADNDT}, \citet{Langanke2001ADNDT}, \citet{Pruet2003ApJS}, and \citet{Suzuki2016ApJ}, as implemented in \texttt{WinNet}. At lower temperatures, these rates are replaced by the experimental decay rates contained in REACLIB.	
	
	In our calculations, neutron star merger ejecta are adopted as the astrophysical environment of interest, as they are particularly favorable for actinide production.
	The ejecta are modeled using parameterized one-zone trajectories, each characterized by the initial electron fraction $Y_{\rm e}$, specific entropy $s$, and expansion timescale $\tau$. All trajectories are initialized at a temperature of $T_0=6$~GK, with the initial composition assumed to be in nuclear statistical equilibrium (NSE). For each trajectory, the initial density $\rho_0$ is determined from the specified $T_0$, $Y_{\rm e}$, and $s$ using the equation of state of \citet{Timmes1999ApJS} under the NSE assumption.
	
	We adopt the density evolution of \citet{Lippuner2015ApJ}. The density initially decreases exponentially and subsequently transitions to the $t^{-3}$ scaling expected for homologous expansion:
	\begin{equation}
		\rho(t) = \rho_0
		\begin{cases}
			\exp\!\left(-t/\tau\right),
			& t\leq 3\tau,\\[3pt]
			\left(\dfrac{3\tau}{et}\right)^3,
			& t>3\tau,
		\end{cases}
	\end{equation}
	where $\tau$ is the expansion timescale and $e$ is Euler's number. The temperature evolution accounts for nuclear heating by updating the entropy in response to changes in the nuclear composition.
	
	We sample $Y_{\rm e}\in[0.005,0.300]$, $s\in[5,40]~k_{\rm B}$ per baryon, and $\tau\in[5,30]$~ms, with grid spacings of $\Delta Y_{\rm e}=0.005$, $\Delta s=5~k_{\rm B}$ per baryon, and $\Delta\tau=5$~ms. This gives a total of $60\times8\times6=2880$ trajectories. For each trajectory, the nuclear reaction network is evolved until active nucleosynthesis has effectively ceased, and the resulting abundance distribution is taken as the initial abundance vector $\boldsymbol{Y}_{\mathrm{ini}}$ for the subsequent radioactive decay calculation described in Section~\ref{sec:radioactive}. The abundances of heavy nuclei with $Z\gtrsim90$ are included to account for their later contributions to the Th and U abundances through decay-chain feeding. For the representative trajectory shown in Figures~\ref{fig:abundance} and \ref{fig:evolve}, we adopt $Y_{\rm e}=0.2$, $s=10~k_{\rm B}$ per baryon, and $\tau=10$~ms, while the full set of trajectories is used to examine variations in the abundance ratios of the relevant progenitor nuclei relative to $^{232}$Th.

	\section{Results}
	\label{sec:result}
	
	\begin{figure*}
		\centering
		\includegraphics[width=0.65\linewidth]{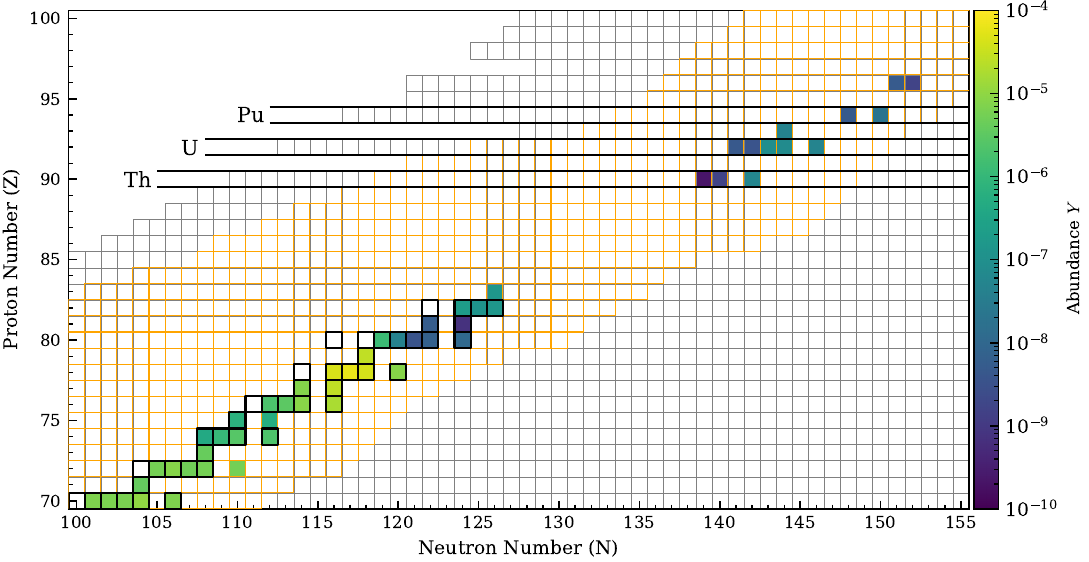}
		\includegraphics[width=0.65\linewidth]{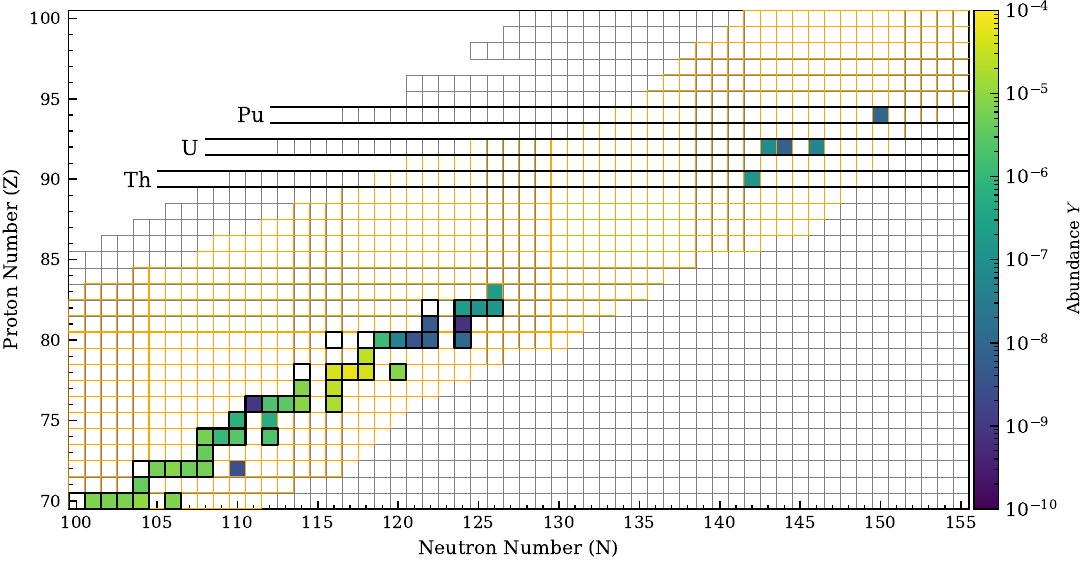}
		\caption{Abundance distributions of heavy nuclei obtained from $r$-process nucleosynthesis simulations at $t=10^6$~yr (top) and $t=10^8$~yr (bottom). The color scale indicates the abundance of each nucleus at the corresponding time. The positions of Th, U, and Pu are marked on the nuclear chart. The calculations adopt a representative neutron star merger trajectory with electron fraction $Y_{\rm e}=0.2$, specific entropy $s=10~k_{\rm B}$ baryon$^{-1}$, and expansion timescale $\tau=10$~ms.}
		\label{fig:abundance}
	\end{figure*}
	
	Figure~\ref{fig:abundance} shows the abundance distributions of heavy nuclei obtained from the $r$-process nucleosynthesis simulations at evolution times of $t=10^6$~yr and $t=10^8$~yr, respectively. 
	For the abundance distribution at $t=10^6$~yr, heavy nuclei in the actinide region remain broadly distributed over the range of $Z\sim90-96$, indicating that a substantial fraction of the heavy nuclei synthesized during the $r$-process has not yet fully decayed. This is because many nuclei in the actinide region possess half-lives longer than $10^6$~yr, allowing them to continuously feed the chronometric nuclei through radioactive decay networks at late times. Among them, several heavy nuclei synthesized in neutron star mergers reach abundances of $\mathcal{O}(10^{-8})$, including $^{232}$Th, $^{235}$U, $^{236}$U, $^{238}$U, and $^{244}$Pu.
	For the $r$-process nucleosynthesis simulations evolved to $t=10^8$~yr, the abundance distribution within the actinide region becomes significantly depleted. The number of surviving nuclei in the $Z\sim90-96$ region decreases substantially, leaving only a small number of long-lived isotopes. These surviving isotopes include $^{232}$Th, $^{235}$U, $^{236}$U, $^{238}$U, and $^{244}$Pu, consistent with the dominant surviving nuclei at $t=10^6$~yr.
	Among these heavy nuclei, the abundance of the chronometric nucleus $^{232}{\rm Th}$ increases by a factor of $\sim2$ from $t=10^6$~yr to $t=10^8$~yr.
	At the same time, the abundance of another chronometric nucleus $^{238}{\rm U}$ remains nearly constant over this time interval. These abundance evolutions suggest that radioactive progenitors continue to feed the chronometric nuclei after freeze-out. Therefore, to clarify the origin of these variations, it is essential to identify the decay pathways that dominate the subsequent evolution of the chronometric nuclei.

	\begin{figure}
		\centering
		\includegraphics[width=1.0\linewidth]{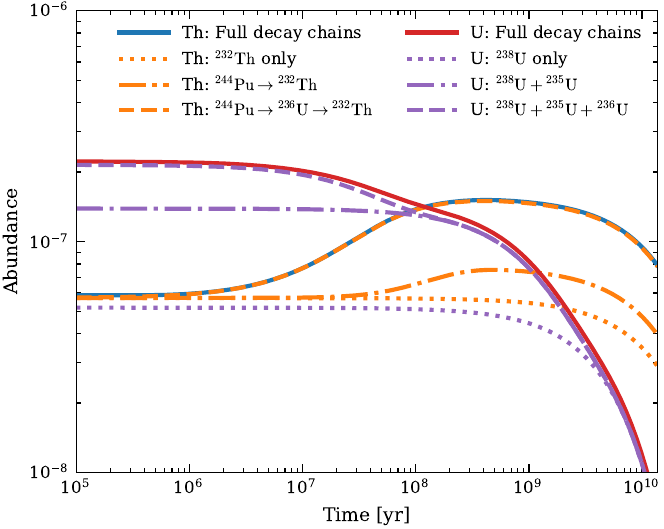}
		\caption{Time evolution of the elemental abundances of Th and U for a representative trajectory with $Y_{\rm e}=0.2$, $s=10~k_{\rm B}$ per baryon, and $\tau=10$~ms. The solid blue and red curves show the results from the full radioactive decay network for Th and U, respectively. For the Th abundance, the orange dotted curve shows the isolated decay of $^{232}$Th, while the orange dash-dotted and dashed curves successively include decay-chain feeding from $^{244}$Pu and $^{236}$U. For the U abundance, the purple dotted curve represents the contribution from $^{238}$U alone, while the purple dash-dotted and dashed curves successively include additional contributions from $^{235}$U and $^{236}$U.}
		\label{fig:evolve}
	\end{figure}
	
	To identify the dominant long-lived progenitor nuclei responsible for decay-chain feeding, we compare the abundance evolution of chronometric nuclei obtained with the full radioactive decay network to that obtained without decay-chain feeding, as shown in Figure~\ref{fig:evolve}.
	For the chronometric element Th, the calculations with and without decay-chain feeding yield nearly identical abundance evolution during the early phase ($t\lesssim10^7$~yr), remaining at $Y_{\rm Th}\sim5.0\times10^{-8}$. However, at $t>10^7$~yr, the Th abundance calculated with decay-chain feeding increases significantly and reaches a peak at $\sim10^9$~yr. The peak abundance of Th is higher than that obtained without decay-chain feeding by a factor of $\sim3$.
	As shown in Figure~\ref{fig:evolve}, when the contributions from heavy nuclei $^{244}$Pu and $^{236}$U are included in the single nucleus decay calculation for Th, the resulting abundance evolution reproduces the results obtained with full radioactive decay network. This suggests that $^{244}$Pu and $^{236}$U are the dominant long-lived progenitor nuclei responsible for the delayed feeding of Th.
	For the other chronometric element U, the contributions from its long-lived isotopes play an important role in shaping its abundance evolution. In the calculation that includes only $^{238}$U, the U abundance remains at $\sim5\times10^{-8}$ and then decreases significantly after $\sim10^9$~yr due to the radioactive decay of $^{238}$U alone. 
	When the long-lived isotopes $^{235}$U and $^{236}$U are taken into account, the U elemental abundance is enhanced relative to the calculation including only $^{238}$U by factors of $\sim2$ and $\sim4$, respectively, and becomes broadly consistent with the result obtained with the full radioactive decay network. 
	This enhancement arises because both $^{235}$U and $^{236}$U have half-lives longer than $10^7$~yr and therefore provide sustained contributions to the U elemental abundance during the early phase. 
	At late times ($t>3\times10^9$~yr), the U elemental abundance is dominated almost entirely by the chronometric nucleus $^{238}$U. 
	This supports the assumption that the U abundance observed in metal-poor stars can be regarded as representative of the $^{238}$U abundance. The abundance evolution of Th and U obtained with our full radioactive decay network is broadly consistent with that reported by \citet{2007PhR...450...97A}. These results suggest that delayed feeding from long-lived progenitor nuclei and long-lived isotopes must be properly accounted for in nuclear chronometry, as it can significantly affect the long-term abundance evolution of chronometric nuclei, particularly for the U/Th chronometer pair.

	The abundance evolution results presented above were obtained from $r$-process nucleosynthesis simulations performed for neutron star merger ejecta as a representative astrophysical site. Numerical relativity simulations have generally shown that the electron fraction of dynamical ejecta from neutron star mergers spans a range of $Y_{\rm e}\simeq0.05-0.25$ \citep{Shibata2019}. In our $r$-process nucleosynthesis simulations, we adopt an electron fraction of $Y_{\rm e}=0.2$, which can be regarded as a conservative choice for actinide production. Actinide synthesis is expected to proceed more efficiently in merger ejecta with lower electron fractions \citep{2022APPSB..32...19W}. For example, the actinide mass fraction obtained for ejecta with $Y_{\rm e}\leq0.1$ reaches $\mathcal{O} (10^{-1}-10^{-2})$, which is about $2-3$ orders of magnitude higher than that obtained for the $Y_{\rm e}=0.2$ case \citep{2022APPSB..32...19W}. This suggests that neutron star mergers are promising astrophysical sites for actinide production and that accounting for delayed feeding from long-lived heavy nuclei is essential and critical for nuclear chronometry.
	
	The abundance evolution shown above indicates that only a small number of long-lived nuclei can continue to modify chronometric abundances at late times. At $t=10^8$~yr, these nuclei include $^{232}$Th, $^{235}$U, $^{236}$U, $^{238}$U, and $^{244}$Pu. Since $^{232}$Th and $^{238}$U are themselves chronometric nuclei, the delayed feeding effects are expected to be dominated primarily by the remaining progenitors $^{235}$U, $^{236}$U, and $^{244}$Pu. Therefore, we further quantify how variations in their initial production ratios affect the derived chronometric ages.
	
\begin{figure*}
		\centering
		\includegraphics[width=1.0\linewidth]{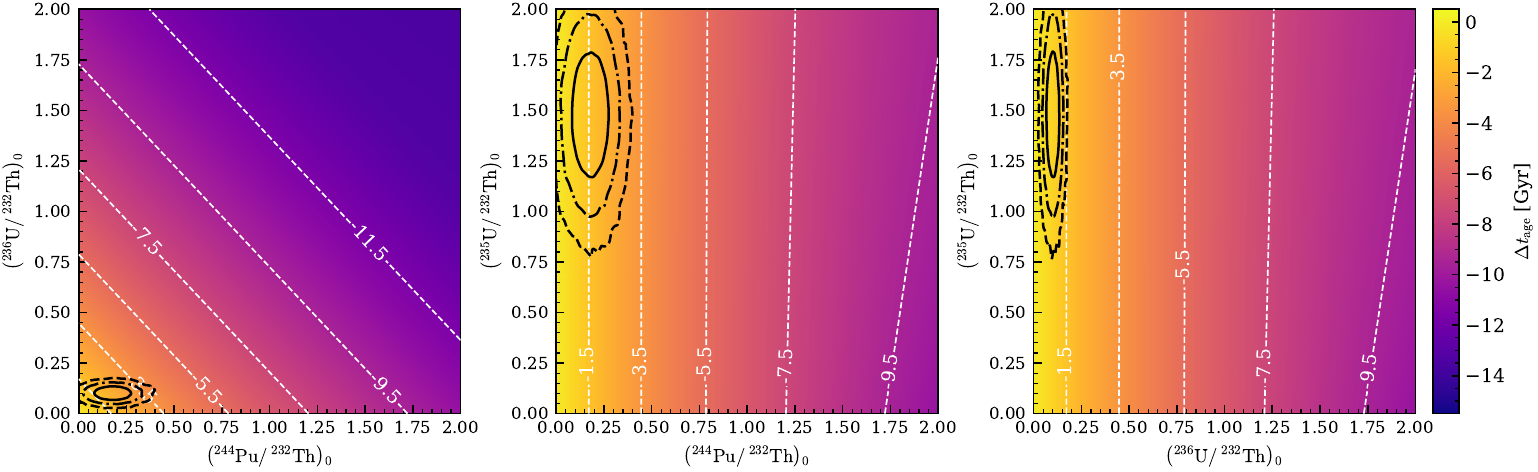}
		\caption{Age correction maps for the U/Th chronometer accounting for decay-chain feeding from long-lived progenitor nuclei. The colour scale represents the age correction $\Delta t_{\mathrm{age}}$ in Gyr, with white dashed contours indicating constant negative values of $\Delta t_{\mathrm{age}}$. The black solid, dash-dotted, and dashed contours indicate the $1\sigma$, $2\sigma$, and $3\sigma$ regions of the abundance ratios derived from the nucleosynthesis simulations, respectively. The initial production ratio $(^{238}\mathrm{U}/^{232}\mathrm{Th})_0$ is normalized such that the classical U/Th chronometer yields an age of $13.8$~Gyr.}
		\label{fig:UTh}
	\end{figure*}

\begin{figure*}
		\centering
		\includegraphics[width=1.0\linewidth]{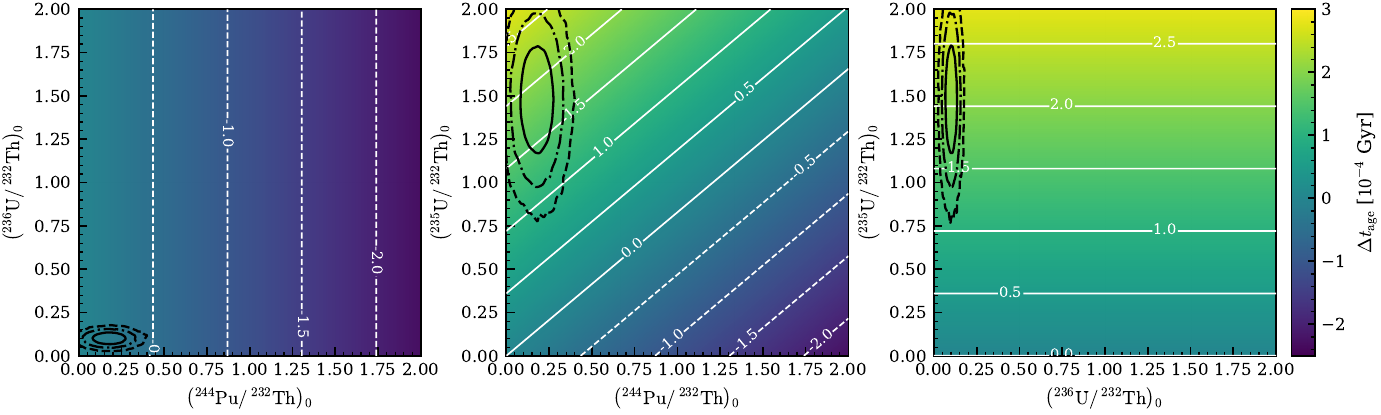}
		\caption{Same as Figure~\ref{fig:UTh}, but for the U/X chronometer, where X denotes a stable $r$-process reference element. The colour scale shows the age correction $\Delta t_{\mathrm{age}}$ in units of $10^{-4}$~Gyr.
		White solid contours indicate zero and positive values of $\Delta t_{\mathrm{age}}$, whereas white dashed contours indicate negative values. The black solid, dash-dotted, and dashed contours indicate the $1\sigma$, $2\sigma$, and $3\sigma$ regions derived from the nucleosynthesis simulations, respectively.}
		\label{fig:UX}
	\end{figure*}

\begin{figure}
		\centering
		\includegraphics[width=1.0\linewidth]{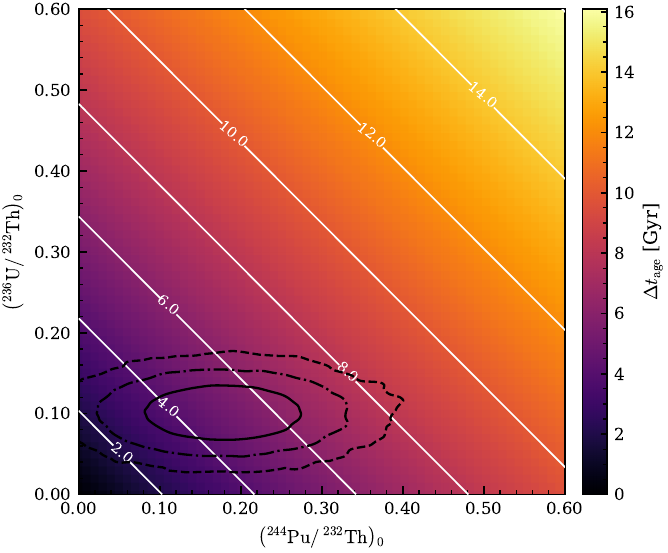}
		\caption{Same as Figure~\ref{fig:UX}, but for the Th/X chronometer. 
		The contributions from the long-lived progenitor nuclei $^{236}$U and $^{244}$Pu are included, and the age correction is shown as a function of their initial abundance ratios relative to $^{232}$Th. The colour scale shows $\Delta t_{\mathrm{age}}$ in Gyr, with white solid contours indicating constant positive values. The black solid, dash-dotted, and dashed contours indicate the $1\sigma$, $2\sigma$, and $3\sigma$ regions derived from the nucleosynthesis simulations, respectively.}
		\label{fig:ThX}
	\end{figure}

	To facilitate a direct comparison of the effects of delayed feeding from long-lived progenitor nuclei on chronometric age estimates, we adopt $^{232}$Th as the reference abundance because it is the longest-lived isotope among the dominant nuclei considered here. The effects of different initial progenitor abundances are then quantified by varying their abundance ratios relative to $^{232}$Th. Specifically, we define the ratios $(^{244}\mathrm{Pu}/^{232}\mathrm{Th})_0$, $(^{235}\mathrm{U}/^{232}\mathrm{Th})_0$, and $(^{236}\mathrm{U}/^{232}\mathrm{Th})_0$ as the initial production ratios immediately after the end of $r$-process nucleosynthesis. Figures~\ref{fig:UTh}-\ref{fig:ThX} present the effects of delayed feeding from long-lived progenitor nuclei on the age estimates derived from the U/Th, U/X, and Th/X chronometer pairs, respectively.
	
	\begin{figure}
		\centering
		\includegraphics[width=1.0\linewidth]{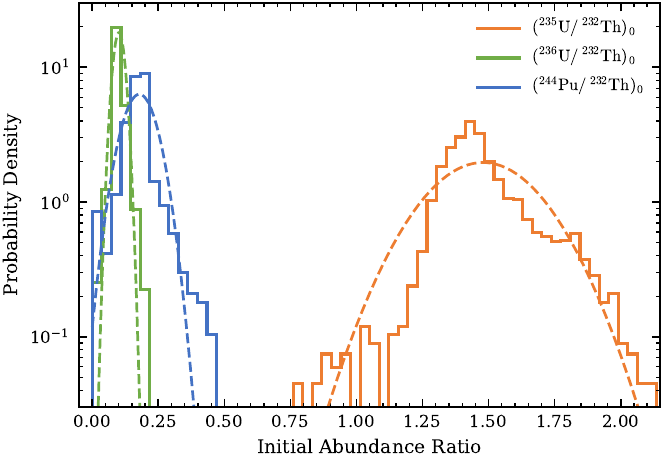}
		\caption{Probability distributions of the abundances ratios of progenitor nuclei relative to $^{232}$Th, obtained from the 2880 one-zone trajectories at $t=10^8$~yr. The orange, green, and blue step curves show the normalized probability distributions of $^{235}$U/$^{232}$Th, $^{236}$U/$^{232}$Th, and $^{244}$Pu/$^{232}$Th, respectively, while the dashed curves show the corresponding Gaussian fits.}
		\label{fig:Ratios}
	\end{figure}

	Figure~\ref{fig:Ratios} shows the distributions of the abundance ratios of the three progenitor nuclei relative to $^{232}$Th obtained from the 2880 $r$-process trajectories. The three ratios exhibit clearly different distributions. The $(^{235}\mathrm{U}/^{232}\mathrm{Th})_0$ ratio has the largest characteristic value, with a relatively broad distribution centred at $\sim1.5$. In contrast, $(^{236}\mathrm{U}/^{232}\mathrm{Th})_0$ is concentrated in a much narrower range around $\sim0.1$, while $(^{244}\mathrm{Pu}/^{232}\mathrm{Th})_0$ shows an intermediate characteristic value of $\sim0.2$ and a broader distrubution than $(^{236}\mathrm{U}/^{232}\mathrm{Th})_0$. The dashed curves show Gaussian fits to the corresponding distributions. These distributions are then used to define the $1\sigma$, $2\sigma$, and $3\sigma$ regions shown in the age correction maps below. These regions represent the range of abundance ratios obtained from the $r$-process nucleosynthesis simulations over the adopted parameter grid.

	Figure~\ref{fig:UTh} shows the age correction maps for the U/Th chronometer when contributions from long-lived progenitor nuclei are included. The resulting correction is negative across the entire explored parameter space, spanning $-15~{\rm Gyr}\lesssim\Delta t_{\rm age}\lesssim0$. Since $\Delta t_{\rm age}=t_{\rm DC}-t_{\rm CL}$, a negative correction means that neglecting delayed decay-chain feeding leads to an overestimate of the U/Th age. The left panel shows that the U/Th age is particularly sensitive to the abundances of $^{236}$U and $^{244}$Pu, with its magnitude increasing systematically as either abundance ratio increases. When the abundance ratio of either $^{236}$U or $^{244}$Pu relative to $^{232}$Th approaches unity, delayed feeding produces an age correction of order $-7$~Gyr. The nearly vertical contours in the middle and right panels indicate that the age correction depends only weakly on $(^{235}\mathrm{U}/^{232}\mathrm{Th})_0$ for fixed values of $(^{244}\mathrm{Pu}/^{232}\mathrm{Th})_0$ or $(^{236}\mathrm{U}/^{232}\mathrm{Th})_0$. The U/Th correction is therefore dominated by delayed feeding from $^{236}$U and $^{244}$Pu. For the range of initial abundances obtained from the nucleosynthesis simulations, the age corrections due to long-lived progenitor nuclei are predominantly within $-3~{\rm Gyr}\lesssim\Delta t_{\rm age}\lesssim0$.

	Figure~\ref{fig:UX} shows the age correction maps for the U/X chronometer. Here the age correction $\Delta t_{\rm age}$ is given in units of $10^{-4}$~Gyr. Across the displayed parameter space, the age correction remains within approximately $-2\times10^{-4}\lesssim\Delta t_{\rm age}\lesssim3\times10^{-4}$~Gyr, which is about four to five orders of magnitude smaller than that for the U/Th chronometer. The left panel shows a weak negative dependence on $(^{244}\mathrm{Pu}/^{232}\mathrm{Th})_0$ and almost no dependence on $(^{236}\mathrm{U}/^{232}\mathrm{Th})_0$. The middle panel shows weak but opposite dependences on $^{244}$Pu and $^{235}$U: the age correction becomes more negative as the $^{244}$Pu ratio increases and more positive as the $^{235}$U ratio increases. The right panel further shows that the variation is driven mainly by $^{235}$U, with only a weak dependence on $^{236}$U. These trends can change the sign of the correction, but its absolute magnitude remains negligible. The U/X chronometer is therefore effectively insensitive to delayed decay-chain feeding over the explored parameter space.
	
	Figure~\ref{fig:ThX} shows the age correction map for the Th/X chronometer. Since the dominant long-lived progenitor nuclei contributing to $^{232}$Th are $^{236}$U and $^{244}$Pu, the correction is shown as a function of their abundance ratios relative to $^{232}$Th. The age correction is positive throughout the explored parameter space and increases with both progenitor ratios, reaching values up to $\sim16$~Gyr. The positive sign arises because delayed feeding increases the late-time $^{232}$Th abundance, so the decay-chain calculation requires a longer time than the classical calculation to reproduce a given observed Th/X ratio. For the range of initial abundances obtained from the nucleosynthesis simulations, the age correction lies within $2~{\rm Gyr}\lesssim\Delta t_{\rm age}\lesssim8~{\rm Gyr}$ and is mainly concentrated around $\sim5$~Gyr. The Th/X chronometer is therefore sensitive to delayed feeding from both $^{236}$U and $^{244}$Pu.

	\section{Conclusion}
	\label{sec:conclusion}
	
	Nuclear cosmochronology provides a means of estimating the ages of astrophysical objects through the radioactive decay of long-lived nuclei synthesized in nucleosynthesis processes. This method compares observed abundances with predicted initial production ratios to infer the time elapsed since nucleosynthesis. However, the dominant uncertainty in nuclear cosmochronology arises from the determination of the initial production ratios. Previous studies have made significant efforts to estimate these ratios using $r$-process nucleosynthesis calculations with different nuclear mass models, nuclear reaction rates, decay properties, and astrophysical conditions \citep{Niu2009PhRvC,Niu2013PhLB,Niu2019PhRvC,Wu2022ApJ}.
	
	In this work, we investigated the impact of decay-chain feeding from long-lived heavy nuclei on nuclear cosmochronology. We construct a self-consistent radioactive decay network incorporating all relevant decay pathways and couple it to $r$-process nucleosynthesis calculations. By tracking the long-term evolution of $r$-process nucleosynthesis based on neutron star merger trajectories, it is found that several actinide nuclei synthesized in the merger ejecta can survive up to $10^8$~yr, and that the Th abundance exhibits a significant enhancement at $\sim10^9$~yr. We analyzed the contribution of each component and identified the dominant long-lived nuclei that shape the evolution of the chronometric elements Th and U, including $^{232}$Th, $^{235}$U, $^{236}$U, $^{238}$U, and $^{244}$Pu. Taking the Th abundance as a fiducial reference, we estimated the age correction for each chronometer by accounting for contributions from these long-lived progenitor nuclei. The U/Th chronometer is sensitive to delayed feeding from the progenitor nuclei $^{236}$U and $^{244}$Pu, which enhances the late-time Th abundance and leads to lower inferred ages than those obtained with the classical U/Th chronometer. For the range of initial abundances obtained from the nucleosynthesis simulations, the age correction can reach approximately $-3$~Gyr. In contrast, the Th/X chronometer shows the opposite behavior, with delayed feeding generally producing positive age corrections because the additional late-time contribution to Th increases the Th/X ratio and therefore leads to older inferred ages relative to the classical approximation. For the Th/X chronometer, the age correction is typically $\sim5$~Gyr within the range of initial abundances obtained from the nucleosynthesis simulations. In addition, the U/X chronometer is largely insensitive to delayed feeding, because the late-time U abundance is dominated by the long-lived isotope $^{238}$U and is only weakly affected by contributions from its progenitor nuclei.

    Our results suggest that the long-term evolution of Th and U can be described by a reduced network of five key isotopes: \(^{232}\mathrm{Th}\), \(^{235}\mathrm{U}\), \(^{236}\mathrm{U}\), \(^{238}\mathrm{U}\), and \(^{244}\mathrm{Pu}\). For the conditions considered here, this reduced network captures the main decay feeding effects on Gyr timescales. However, the relevant progenitor nuclei may depend on the nuclear physics inputs and astrophysical conditions, so full decay-chain calculations remain necessary to assess the applicability of such a simplification.
	
	We note that neutron star mergers are among the most promising sites for actinide production \citep{2022APPSB..32...19W}. 
	In this work, we adopt \(Y_{\rm e}=0.2\) as a representative case to illustrate the evolution of actinide abundances in merger ejecta. However, neutron star mergers can also produce more neutron-rich ejecta with lower \(Y_{\rm e}\), which favor the synthesis of a wider range of actinide nuclei and can lead to substantially larger actinide abundances. For ejecta with \(Y_{\rm e}=0.1\), the actinide mass fraction can reach \(\mathcal{O}(10^{-2}\text{--}10^{-1})\), about \(2\text{--}3\) orders of magnitude higher than that for ejecta with \(Y_{\rm e}=0.2\)  \citep{2022APPSB..32...19W}. This suggests that delayed feeding from long-lived heavy nuclei may become particularly important in actinide-rich ejecta and should be taken into account when applying the U/Th and Th/X chronometers.
	
	Neutron star mergers are not the only possible astrophysical sites of actinide production \citep{Chen2024MNRAS,2026SCPMA..6909513Q}. Other \(r\)-process environments, such as magneto-rotational supernovae, may also synthesize heavy \(r\)-process nuclei and actinides under sufficiently neutron-rich conditions \citep[e.g.][]{Cescutti2014AAP,Kobayashi2023ApJL,Lian2023MNRAS}. Different astrophysical sites can produce different initial abundances of Th, U, and their long-lived progenitors, leading to corresponding variations in decay-chain feeding and the inferred chronometric ages. The age corrections obtained here therefore apply to the neutron star merger conditions considered in this work and should not be regarded as universal. However, the decay-chain formalism is independent of the astrophysical site and can be directly applied to abundance distributions predicted for other $r$-process environments.

	\section*{Acknowledgements}
	This work is supported by the National Key R\&D Program of China (2024YFA1611704), the Guangxi Science and Technology Innovation Platform Program (Guike LT2600640026), and the National Natural Science Foundation of China (Grant Nos. 12403043, 12133003, and 12473038). MHC also acknowledges the support from the Guangxi Bagui Young Scholars Foundation. EWL is also supported by the Guangxi Talent Program (Highland of Innovation Talents).

	
    \section*{Data Availability}
	
    The data underlying this article will be shared upon reasonable request to the corresponding author.

	
	
	\bibliographystyle{mnras}
	\bibliography{references}

@ARTICLE{Reichert2023ApJS,
       author = {{Reichert}, M. and {Winteler}, C. and {Korobkin}, O. and {Arcones}, A. and {Bliss}, J. and {Eichler}, M. and {Frischknecht}, U. and {Fr{\"o}hlich}, C. and {Hirschi}, R. and {Jacobi}, M. and et al.},
        title = "{The Nuclear Reaction Network WinNet}",
      journal = {\apjs},
         year = 2023,
        month = oct,
       volume = {268},
       number = {2},
          eid = {66},
        pages = {66},
          doi = {10.3847/1538-4365/acf033},
archivePrefix = {arXiv},
       eprint = {2305.07048},
 primaryClass = {astro-ph.IM},
       adsurl = {https://ui.adsabs.harvard.edu/abs/2023ApJS..268...66R}
}

@ARTICLE{Timmes1999ApJS,
       author = {{Timmes}, F.~X. and {Arnett}, Dave},
        title = "{The Accuracy, Consistency, and Speed of Five Equations of State for Stellar Hydrodynamics}",
      journal = {\apjs},
         year = 1999,
        month = nov,
       volume = {125},
       number = {1},
        pages = {277-294},
          doi = {10.1086/313271},
       adsurl = {https://ui.adsabs.harvard.edu/abs/1999ApJS..125..277T}
}

@ARTICLE{Lippuner2015ApJ,
       author = {{Lippuner}, Jonas and {Roberts}, Luke F.},
        title = "{r-process Lanthanide Production and Heating Rates in Kilonovae}",
      journal = {\apj},
         year = 2015,
        month = dec,
       volume = {815},
       number = {2},
          eid = {82},
        pages = {82},
          doi = {10.1088/0004-637X/815/2/82},
archivePrefix = {arXiv},
       eprint = {1508.03133},
 primaryClass = {astro-ph.HE},
       adsurl = {https://ui.adsabs.harvard.edu/abs/2015ApJ...815...82L}
}

@ARTICLE{Koning2023EPJA,
       author = {{Koning}, Arjan and {Hilaire}, Stephane and {Goriely}, Stephane},
        title = "{TALYS: modeling of nuclear reactions}",
      journal = {European Physical Journal A},
         year = 2023,
        month = jun,
       volume = {59},
       number = {6},
          eid = {131},
        pages = {131},
          doi = {10.1140/epja/s10050-023-01034-3},
       adsurl = {https://ui.adsabs.harvard.edu/abs/2023EPJA...59..131K}
}

@ARTICLE{Wang2021CPC,
       author = {{Wang}, Meng and {Huang}, W.~J. and {Kondev}, F.~G. and {Audi}, G. and {Naimi}, S.},
        title = "{The AME 2020 atomic mass evaluation (II). Tables, graphs and references}",
      journal = {Chinese Physics C},
         year = 2021,
        month = mar,
       volume = {45},
       number = {3},
          eid = {030003},
        pages = {030003},
          doi = {10.1088/1674-1137/abddaf},
       adsurl = {https://ui.adsabs.harvard.edu/abs/2021ChPhC..45c0003W}
}

@ARTICLE{Kondev2021CPC,
       author = {{Kondev}, F.~G. and {Wang}, M. and {Huang}, W.~J. and {Naimi}, S. and {Audi}, G.},
        title = "{The NUBASE2020 evaluation of nuclear physics properties}",
      journal = {Chinese Physics C},
         year = 2021,
        month = mar,
       volume = {45},
       number = {3},
          eid = {030001},
        pages = {030001},
          doi = {10.1088/1674-1137/abddae},
       adsurl = {https://ui.adsabs.harvard.edu/abs/2021ChPhC..45c0001K}
}

@ARTICLE{Cyburt2010ApJS,
       author = {{Cyburt}, Richard H. and {Amthor}, A. Matthew and {Ferguson}, Ryan and {Meisel}, Zach and {Smith}, Karl and {Warren}, Scott and {Heger}, Alexander and {Hoffman}, R.~D. and {Rauscher}, Thomas and {Sakharuk}, Alexander and {Schatz}, Hendrik and {Thielemann}, F.~K. and {Wiescher}, Michael},
        title = "{The JINA REACLIB Database: Its Recent Updates and Impact on Type-I X-ray Bursts}",
      journal = {\apjs},
         year = 2010,
        month = jul,
       volume = {189},
       number = {1},
        pages = {240-252},
          doi = {10.1088/0067-0049/189/1/240},
       adsurl = {https://ui.adsabs.harvard.edu/abs/2010ApJS..189..240C}
}

@ARTICLE{Moller2016ADNDT,
       author = {{M{\"o}ller}, P. and {Sierk}, A.~J. and {Ichikawa}, T. and {Sagawa}, H.},
        title = "{Nuclear ground-state masses and deformations: FRDM(2012)}",
      journal = {Atomic Data and Nuclear Data Tables},
         year = 2016,
        month = may,
       volume = {109},
        pages = {1-204},
          doi = {10.1016/j.adt.2015.10.002},
archivePrefix = {arXiv},
       eprint = {1508.06294},
 primaryClass = {nucl-th},
       adsurl = {https://ui.adsabs.harvard.edu/abs/2016ADNDT.109....1M}
}

@ARTICLE{Moller2015PRC,
       author = {{M{\"o}ller}, Peter and {Sierk}, Arnold J. and {Ichikawa}, Takatoshi and {Iwamoto}, Akira and {Mumpower}, Matthew},
        title = "{Fission barriers at the end of the chart of the nuclides}",
      journal = {\prc},
         year = 2015,
        month = feb,
       volume = {91},
       number = {2},
          eid = {024310},
        pages = {024310},
          doi = {10.1103/PhysRevC.91.024310},
       adsurl = {https://ui.adsabs.harvard.edu/abs/2015PhRvC..91b4310M}
}

@ARTICLE{Panov2005NuPhA,
       author = {{Panov}, I.~V. and {Kolbe}, E. and {Pfeiffer}, B. and {Rauscher}, T. and {Kratz}, K.-L. and {Thielemann}, F.-K.},
        title = "{Calculations of fission rates for r-process nucleosynthesis}",
      journal = {\nphysa},
         year = 2005,
        month = jan,
       volume = {747},
       number = {2-4},
        pages = {633-654},
          doi = {10.1016/j.nuclphysa.2004.09.115},
archivePrefix = {arXiv},
       eprint = {astro-ph/0412654},
 primaryClass = {astro-ph},
       adsurl = {https://ui.adsabs.harvard.edu/abs/2005NuPhA.747..633P}
}

@ARTICLE{Panov2010A&A,
       author = {{Panov}, I.~V. and {Korneev}, I. Yu. and {Rauscher}, T. and {Mart{\'\i}nez-Pinedo}, G. and {Keli{\'c}-Heil}, A. and {Zinner}, N.~T. and {Thielemann}, F.-K.},
        title = "{Neutron-induced astrophysical reaction rates for translead nuclei}",
      journal = {\aap},
         year = 2010,
        month = apr,
       volume = {513},
          eid = {A61},
        pages = {A61},
          doi = {10.1051/0004-6361/200911967},
archivePrefix = {arXiv},
       eprint = {0911.2181},
 primaryClass = {astro-ph.SR},
       adsurl = {https://ui.adsabs.harvard.edu/abs/2010A&A...513A..61P}
}

@ARTICLE{Khuyagbaatar2020NuPhA,
       author = {{Khuyagbaatar}, J.},
        title = "{Spontaneous fission half-lives of the heaviest nuclei: Semi-empirical predictions}",
      journal = {\nphysa},
         year = 2020,
        month = oct,
       volume = {1002},
          eid = {121958},
        pages = {121958},
          doi = {10.1016/j.nuclphysa.2020.121958},
       adsurl = {https://ui.adsabs.harvard.edu/abs/2020NuPhA100221958K}
}

@ARTICLE{Kodama1975NPA,
       author = {{Kodama}, Takeshi and {Takahashi}, Kohji},
        title = "{R-process nucleosynthesis and nuclei far from the region of {\ensuremath{\beta}}-stability}",
      journal = {\nphysa},
         year = 1975,
        month = mar,
       volume = {239},
       number = {3},
        pages = {489-510},
          doi = {10.1016/0375-9474(75)90381-4},
       adsurl = {https://ui.adsabs.harvard.edu/abs/1975NuPhA.239..489K}
}

@ARTICLE{Moller2019ADNDT,
       author = {{M{\"o}ller}, P. and {Mumpower}, M.~R. and {Kawano}, T. and {Myers}, W.~D.},
        title = "{Nuclear properties for astrophysical and radioactive-ion-beam applications (II)}",
      journal = {Atomic Data and Nuclear Data Tables},
         year = 2019,
        month = jan,
       volume = {125},
        pages = {1-192},
          doi = {10.1016/j.adt.2018.03.003},
       adsurl = {https://ui.adsabs.harvard.edu/abs/2019ADNDT.125....1M}
}

@ARTICLE{Fuller1985ApJ,
       author = {{Fuller}, G.~M. and {Fowler}, W.~A. and {Newman}, M.~J.},
        title = "{Stellar weak interaction rates for intermediate-mass nuclei. IV - Interpolation procedures for rapidly varying lepton capture rates using effective log (ft)-values}",
      journal = {\apj},
         year = 1985,
        month = jun,
       volume = {293},
        pages = {1-16},
          doi = {10.1086/163208},
       adsurl = {https://ui.adsabs.harvard.edu/abs/1985ApJ...293....1F}
}

@ARTICLE{Oda1994ADNDT,
       author = {{Oda}, T. and {Hino}, M. and {Muto}, K. and {Takahara}, M. and {Sato}, K.},
        title = "{Rate Tables for the Weak Processes of sd-Shell Nuclei in Stellar Matter}",
      journal = {Atomic Data and Nuclear Data Tables},
         year = 1994,
        month = mar,
       volume = {56},
       number = {2},
        pages = {231-403},
          doi = {10.1006/adnd.1994.1007},
       adsurl = {https://ui.adsabs.harvard.edu/abs/1994ADNDT..56..231O}
}

@ARTICLE{Langanke2001ADNDT,
       author = {{Langanke}, K. and {Mart{\'\i}nez-Pinedo}, G.},
        title = "{Rate Tables for the Weak Processes of pf-SHELL Nuclei in Stellar Environments}",
      journal = {Atomic Data and Nuclear Data Tables},
         year = 2001,
        month = sep,
       volume = {79},
       number = {1},
        pages = {1-46},
          doi = {10.1006/adnd.2001.0865},
       adsurl = {https://ui.adsabs.harvard.edu/abs/2001ADNDT..79....1L}
}

@ARTICLE{Pruet2003ApJS,
       author = {{Pruet}, Jason and {Fuller}, George M.},
        title = "{Estimates of Stellar Weak Interaction Rates for Nuclei in the Mass Range A=65-80}",
      journal = {\apjs},
         year = 2003,
        month = nov,
       volume = {149},
       number = {1},
        pages = {189-203},
          doi = {10.1086/376753},
archivePrefix = {arXiv},
       eprint = {astro-ph/0211262},
 primaryClass = {astro-ph},
       adsurl = {https://ui.adsabs.harvard.edu/abs/2003ApJS..149..189P}
}

@ARTICLE{Suzuki2016ApJ,
       author = {{Suzuki}, Toshio and {Toki}, Hiroshi and {Nomoto}, Ken'ichi},
        title = "{Electron-capture and {\ensuremath{\beta}}-decay Rates for sd-Shell Nuclei in Stellar Environments Relevant to High-density O-Ne-Mg Cores}",
      journal = {\apj},
         year = 2016,
        month = feb,
       volume = {817},
       number = {2},
          eid = {163},
        pages = {163},
          doi = {10.3847/0004-637X/817/2/163},
archivePrefix = {arXiv},
       eprint = {1512.00132},
 primaryClass = {astro-ph.SR},
       adsurl = {https://ui.adsabs.harvard.edu/abs/2016ApJ...817..163S}
}

@ARTICLE{Nobre2025ENDF,
       author = {{Nobre}, G.~P.~A. and {Capote}, R. and {Pigni}, M.~T. and {Trkov}, A. and {Mattoon}, C.~M. and {Neudecker}, D. and {Brown}, D.~A. and {Chadwick}, M.~B. and {Kahler}, A.~C. and {Kleedtke}, N.~A. and {Zerkle}, M. and {Hawari}, A.~I. and {Chapman}, C.~W. and {Fleming}, N.~C. and {Wormald}, J.~L. and {Rami{\'c}}, K. and {Danon}, Y. and {Gibson}, N.~A. and {Brain}, P. and {Paris}, M.~W. and {Hale}, G.~M. and {Thompson}, I.~J. and {Barry}, D.~P. and {Stetcu}, I. and {Haeck}, W. and {Lovell}, A.~E. and {Mumpower}, M.~R. and {Potel}, G. and {Kravvaris}, K. and {Noguere}, G. and {McDonnell}, J.~D. and {Carlson}, A.~D. and {Dunn}, M. and {Kawano}, T. and {Wiarda}, D. and {Al-Qasir}, I. and {Arbanas}, G. and {Arcilla}, R. and {Beck}, B. and {Bernard}, D. and {Beyer}, R. and {Brown}, J.~M. and {Cabellos}, O. and {Casperson}, R.~J. and {Cheng}, Y. and {Chimanski}, E.~V. and {Coles}, R. and {Cornock}, M. and {Cotchen}, J. and {Crozier}, J.~P.~W. and {Cullen}, D.~E. and {Daskalakis}, A. and {Descalle}, M.-A. and {DiJulio}, D.~D. and {Dimitriou}, P. and {Dreyfuss}, A.~C. and {Dur{\'a}n}, I. and {Ferrer}, R. and {Gaines}, T. and {Gillette}, V. and {Gert}, G. and {Guber}, K.~H. and {Haverkamp}, J.~D. and {Herman}, M.~W. and {Holmes}, J. and {Hursin}, M. and {Jisrawi}, N. and {Junghans}, A.~R. and {Kelly}, K.~J. and {Kim}, H.~I. and {Kim}, K.~S. and {Koning}, A.~J. and {Ko{\v{s}}t{\'a}l}, M. and {Laramee}, B.~K. and {Lauer-Coles}, A. and {Leal}, L. and {Lee}, H.~Y. and {Lewis}, A.~M. and {Malec}, J. and {M{\'a}rquez Dami{\'a}n}, J.~I. and {Marshall}, W.~J. and {Mattera}, A. and {Muhrer}, G. and {Ney}, A. and {Ormand}, W.~E. and {Parsons}, D.~K. and {Percher}, C.~M. and {Pronyaev}, V.~G. and {Qteish}, A. and {Quaglioni}, S. and {Rapp}, M. and {Ressler}, J.~J. and {Rising}, M. and {Rochman}, D. and {Romano}, P.~K. and {Roubtsov}, D. and {Schnabel}, G. and {Schulc}, M. and {Siemers}, G.~J. and {Sonzogni}, A.~A. and {Talou}, P. and {Thompson}, J. and {Trumbull}, T.~H. and {van der Marck}, S.~C. and {Vorabbi}, M. and {Wemple}, C. and {Wendt}, K.~A. and {White}, M. and {Wright}, R.~Q.},
        title = "{ENDF/B-VIII.1: Updated Nuclear Reaction Data Library for Science and Applications}",
      journal = {arXiv e-prints},
         year = 2025,
        month = nov,
          eid = {arXiv:2511.03564},
        pages = {arXiv:2511.03564},
          doi = {10.48550/arXiv.2511.03564},
archivePrefix = {arXiv},
       eprint = {2511.03564},
 primaryClass = {physics.app-ph},
       adsurl = {https://ui.adsabs.harvard.edu/abs/2025arXiv251103564N}
}

@ARTICLE{Burbidge1957RvMP,
       author = {{Burbidge}, E. Margaret and {Burbidge}, G.~R. and {Fowler}, William A. and {Hoyle}, F.},
        title = "{Synthesis of the Elements in Stars}",
      journal = {Reviews of Modern Physics},
         year = 1957,
        month = oct,
       volume = {29},
       number = {4},
        pages = {547-650},
          doi = {10.1103/RevModPhys.29.547},
       adsurl = {https://ui.adsabs.harvard.edu/abs/1957RvMP...29..547B}
}

@ARTICLE{Kajino2019,
       author = {{Kajino}, T. and {Aoki}, W. and {Balantekin}, A.~B. and {Diehl}, R. and {Famiano}, M.~A. and {Mathews}, G.~J.},
        title = "{Current status of r-process nucleosynthesis}",
      journal = {Progress in Particle and Nuclear Physics},
         year = 2019,
        month = jul,
       volume = {107},
        pages = {109-166},
          doi = {10.1016/j.ppnp.2019.02.008},
archivePrefix = {arXiv},
       eprint = {1906.05002},
 primaryClass = {astro-ph.HE},
       adsurl = {https://ui.adsabs.harvard.edu/abs/2019PrPNP.107..109K}
}

@ARTICLE{Cowan2021RvMP,
       author = {{Cowan}, John J. and {Sneden}, Christopher and {Lawler}, James E. and {Aprahamian}, Ani and {Wiescher}, Michael and {Langanke}, Karlheinz and {Mart{\'\i}nez-Pinedo}, Gabriel and {Thielemann}, Friedrich-Karl},
        title = "{Origin of the heaviest elements: The rapid neutron-capture process}",
      journal = {Reviews of Modern Physics},
         year = 2021,
        month = jan,
       volume = {93},
       number = {1},
          eid = {015002},
        pages = {015002},
          doi = {10.1103/RevModPhys.93.015002},
archivePrefix = {arXiv},
       eprint = {1901.01410},
 primaryClass = {astro-ph.HE},
       adsurl = {https://ui.adsabs.harvard.edu/abs/2021RvMP...93a5002C}
}

@ARTICLE{Shibata2019,
       author = {{Shibata}, Masaru and {Hotokezaka}, Kenta},
        title = "{Merger and Mass Ejection of Neutron Star Binaries}",
      journal = {Annual Review of Nuclear and Particle Science},
         year = 2019,
        month = oct,
       volume = {69},
        pages = {41-64},
          doi = {10.1146/annurev-nucl-101918-023625},
archivePrefix = {arXiv},
       eprint = {1908.02350},
 primaryClass = {astro-ph.HE},
       adsurl = {https://ui.adsabs.harvard.edu/abs/2019ARNPS..69...41S}
}

@ARTICLE{Kasen2017Nature,
       author = {{Kasen}, Daniel and {Metzger}, Brian and {Barnes}, Jennifer and {Quataert}, Eliot and {Ramirez-Ruiz}, Enrico},
        title = "{Origin of the heavy elements in binary neutron-star mergers from a gravitational-wave event}",
      journal = {\nat},
         year = 2017,
        month = nov,
       volume = {551},
       number = {7678},
        pages = {80-84},
          doi = {10.1038/nature24453},
archivePrefix = {arXiv},
       eprint = {1710.05463},
 primaryClass = {astro-ph.HE},
       adsurl = {https://ui.adsabs.harvard.edu/abs/2017Natur.551...80K}
}

@ARTICLE{Fowler1960,
       author = {{Fowler}, William A. and {Hoyle}, F.},
        title = "{Nuclear Cosmochronology.}",
      journal = {\aj},
         year = 1960,
        month = jan,
       volume = {65},
        pages = {345},
          doi = {10.1086/108117},
       adsurl = {https://ui.adsabs.harvard.edu/abs/1960AJ.....65R.345F}
}

@ARTICLE{Schramm1970ApJ,
       author = {{Schramm}, David N. and {Wasserburg}, G.~J.},
        title = "{Nucleochronologies and the Mean Age of the Elements}",
      journal = {\apj},
         year = 1970,
        month = oct,
       volume = {162},
        pages = {57},
          doi = {10.1086/150634},
       adsurl = {https://ui.adsabs.harvard.edu/abs/1970ApJ...162...57S}
}

@ARTICLE{Wene1975A&A,
       author = {{Wene}, C.-O.},
        title = "{The effect of delayed fission in nucleo-cosmochronology.}",
      journal = {\aap},
         year = 1975,
        month = nov,
       volume = {44},
       number = {1},
        pages = {233-236},
       adsurl = {https://ui.adsabs.harvard.edu/abs/1975A&A....44..233W}
}

@ARTICLE{Thielemann1983A&A,
       author = {{Thielemann}, F.-K. and {Metzinger}, J. and {Klapdor}, H.~V.},
        title = "{New actinide chronometer production ratios and the age of the Galaxy}",
      journal = {\aap},
         year = 1983,
        month = jun,
       volume = {123},
       number = {1},
        pages = {162-169},
       adsurl = {https://ui.adsabs.harvard.edu/abs/1983A&A...123..162T}
}

@ARTICLE{Meyer1986,
       author = {{Meyer}, Bradley S. and {Schramm}, David N.},
        title = "{General Constraints on the Age and Chemical Evolution of the Galaxy}",
      journal = {\apj},
         year = 1986,
        month = dec,
       volume = {311},
        pages = {406},
          doi = {10.1086/164781},
       adsurl = {https://ui.adsabs.harvard.edu/abs/1986ApJ...311..406M}
}

@ARTICLE{Cowan1987ApJ,
       author = {{Cowan}, John J. and {Thielemann}, F.-K. and {Truran}, J.~W.},
        title = "{Nuclear Chronometers from the r-Process and the Age of the Galaxy}",
      journal = {\apj},
         year = 1987,
        month = dec,
       volume = {323},
        pages = {543},
          doi = {10.1086/165850},
       adsurl = {https://ui.adsabs.harvard.edu/abs/1987ApJ...323..543C}
}

@ARTICLE{Cowan1991PhR,
       author = {{Cowan}, John J. and {Thielemann}, Friedrich-Karl and {Truran}, James W.},
        title = "{The R-process and nucleochronology}",
      journal = {\physrep},
         year = 1991,
        month = nov,
       volume = {208},
       number = {4-5},
        pages = {267-394},
          doi = {10.1016/0370-1573(91)90070-3},
       adsurl = {https://ui.adsabs.harvard.edu/abs/1991PhR...208..267C}
}

@ARTICLE{Niu2009PhRvC,
       author = {{Niu}, Zhongming and {Sun}, Baohua and {Meng}, Jie},
        title = "{Influence of nuclear physics inputs and astrophysical conditions on the Th/U chronometer}",
      journal = {\prc},
         year = 2009,
        month = dec,
       volume = {80},
       number = {6},
          eid = {065806},
        pages = {065806},
          doi = {10.1103/PhysRevC.80.065806},
archivePrefix = {arXiv},
       eprint = {0912.1669},
 primaryClass = {nucl-th},
       adsurl = {https://ui.adsabs.harvard.edu/abs/2009PhRvC..80f5806N}
}

@ARTICLE{Niu2013PhLB,
       author = {{Niu}, Z.~M. and {Niu}, Y.~F. and {Liang}, H.~Z. and {Long}, W.~H. and {Nik{\v{s}}i{\'c}}, T. and {Vretenar}, D. and {Meng}, J.},
        title = "{{\ensuremath{\beta}}-decay half-lives of neutron-rich nuclei and matter flow in the r-process}",
      journal = {Physics Letters B},
         year = 2013,
        month = jun,
       volume = {723},
       number = {1-3},
        pages = {172-176},
          doi = {10.1016/j.physletb.2013.04.048},
archivePrefix = {arXiv},
       eprint = {1210.0680},
 primaryClass = {nucl-th},
       adsurl = {https://ui.adsabs.harvard.edu/abs/2013PhLB..723..172N}
}

@ARTICLE{Niu2019PhRvC,
       author = {{Niu}, Z.~M. and {Liang}, H.~Z. and {Sun}, B.~H. and {Long}, W.~H. and {Niu}, Y.~F.},
        title = "{Predictions of nuclear {\ensuremath{\beta}} -decay half-lives with machine learning and their impact on r -process nucleosynthesis}",
      journal = {\prc},
         year = 2019,
        month = jun,
       volume = {99},
       number = {6},
          eid = {064307},
        pages = {064307},
          doi = {10.1103/PhysRevC.99.064307},
archivePrefix = {arXiv},
       eprint = {1810.03156},
 primaryClass = {nucl-th},
       adsurl = {https://ui.adsabs.harvard.edu/abs/2019PhRvC..99f4307N}
}

@ARTICLE{Wu2022ApJ,
       author = {{Wu}, X.~H. and {Zhao}, P.~W. and {Zhang}, S.~Q. and {Meng}, J.},
        title = "{High-precision Nuclear Chronometer for the Cosmos}",
      journal = {\apj},
         year = 2022,
        month = dec,
       volume = {941},
       number = {2},
          eid = {152},
        pages = {152},
          doi = {10.3847/1538-4357/aca526},
archivePrefix = {arXiv},
       eprint = {2108.06104},
 primaryClass = {nucl-th},
       adsurl = {https://ui.adsabs.harvard.edu/abs/2022ApJ...941..152W}
}

@ARTICLE{Cowan1999ApJ,
       author = {{Cowan}, John J. and {Pfeiffer}, B. and {Kratz}, K.-L. and {Thielemann}, F.-K. and {Sneden}, Christopher and {Burles}, Scott and {Tytler}, David and {Beers}, Timothy C.},
        title = "{R-Process Abundances and Chronometers in Metal-poor Stars}",
      journal = {\apj},
         year = 1999,
        month = aug,
       volume = {521},
       number = {1},
        pages = {194-205},
          doi = {10.1086/307512},
archivePrefix = {arXiv},
       eprint = {astro-ph/9808272},
 primaryClass = {astro-ph},
       adsurl = {https://ui.adsabs.harvard.edu/abs/1999ApJ...521..194C}
}

@ARTICLE{MP:Hill2002,
       author = {{Hill}, V. and {Plez}, B. and {Cayrel}, R. and {Beers}, T.~C. and {Nordstr{\"o}m}, B. and {Andersen}, J. and {Spite}, M. and {Spite}, F. and {Barbuy}, B. and {Bonifacio}, P. and {Depagne}, E. and {Fran{\c{c}}ois}, P. and {Primas}, F.},
        title = "{First stars. I. The extreme r-element rich, iron-poor halo giant CS 31082-001. Implications for the r-process site(s) and radioactive cosmochronology}",
      journal = {\aap},
         year = 2002,
        month = may,
       volume = {387},
        pages = {560-579},
          doi = {10.1051/0004-6361:20020434},
archivePrefix = {arXiv},
       eprint = {astro-ph/0203462},
 primaryClass = {astro-ph},
       adsurl = {https://ui.adsabs.harvard.edu/abs/2002A&A...387..560H}
}

@ARTICLE{MP:Cowan2002,
       author = {{Cowan}, John J. and {Sneden}, Christopher and {Burles}, Scott and {Ivans}, Inese I. and {Beers}, Timothy C. and {Truran}, James W. and {Lawler}, James E. and {Primas}, Francesca and {Fuller}, George M. and {Pfeiffer}, Bernd and {Kratz}, Karl-Ludwig},
        title = "{The Chemical Composition and Age of the Metal-poor Halo Star BD +17{\textdegree}3248}",
      journal = {\apj},
         year = 2002,
        month = jun,
       volume = {572},
       number = {2},
        pages = {861-879},
          doi = {10.1086/340347},
archivePrefix = {arXiv},
       eprint = {astro-ph/0202429},
 primaryClass = {astro-ph},
       adsurl = {https://ui.adsabs.harvard.edu/abs/2002ApJ...572..861C}
}

@ARTICLE{MP:Frebel2007,
       author = {{Frebel}, Anna and {Christlieb}, Norbert and {Norris}, John E. and {Thom}, Christopher and {Beers}, Timothy C. and {Rhee}, Jaehyon},
        title = "{Discovery of HE 1523-0901, a Strongly r-Process-enhanced Metal-poor Star with Detected Uranium}",
      journal = {\apjl},
         year = 2007,
        month = may,
       volume = {660},
       number = {2},
        pages = {L117-L120},
          doi = {10.1086/518122},
archivePrefix = {arXiv},
       eprint = {astro-ph/0703414},
 primaryClass = {astro-ph},
       adsurl = {https://ui.adsabs.harvard.edu/abs/2007ApJ...660L.117F}
}

@ARTICLE{MP:Hill2017,
       author = {{Hill}, V. and {Christlieb}, N. and {Beers}, T.~C. and {Barklem}, P.~S. and {Kratz}, K.-L. and {Nordstr{\"o}m}, B. and {Pfeiffer}, B. and {Farouqi}, K.},
        title = "{The Hamburg/ESO R-process Enhanced Star survey (HERES). XI. The highly r-process-enhanced star CS 29497-004}",
      journal = {\aap},
         year = 2017,
        month = nov,
       volume = {607},
          eid = {A91},
        pages = {A91},
          doi = {10.1051/0004-6361/201629092},
archivePrefix = {arXiv},
       eprint = {1608.07463},
 primaryClass = {astro-ph.GA},
       adsurl = {https://ui.adsabs.harvard.edu/abs/2017A&A...607A..91H}
}

@ARTICLE{MP:Placco2017,
       author = {{Placco}, Vinicius M. and {Holmbeck}, Erika M. and {Frebel}, Anna and {Beers}, Timothy C. and {Surman}, Rebecca A. and {Ji}, Alexander P. and {Ezzeddine}, Rana and {Points}, Sean D. and {Kaleida}, Catherine C. and {Hansen}, Terese T. and {Sakari}, Charli M. and {Casey}, Andrew R.},
        title = "{RAVE J203843.2-002333: The First Highly R-process-enhanced Star Identified in the RAVE Survey}",
      journal = {\apj},
         year = 2017,
        month = jul,
       volume = {844},
       number = {1},
          eid = {18},
        pages = {18},
          doi = {10.3847/1538-4357/aa78ef},
archivePrefix = {arXiv},
       eprint = {1706.02934},
 primaryClass = {astro-ph.SR},
       adsurl = {https://ui.adsabs.harvard.edu/abs/2017ApJ...844...18P}
}

@ARTICLE{MP:Holmbeck2018,
       author = {{Holmbeck}, Erika M. and {Beers}, Timothy C. and {Roederer}, Ian U. and {Placco}, Vinicius M. and {Hansen}, Terese T. and {Sakari}, Charli M. and {Sneden}, Christopher and {Liu}, Chao and {Lee}, Young Sun and {Cowan}, John J. and {Frebel}, Anna},
        title = "{The R-Process Alliance: 2MASS J09544277+5246414, the Most Actinide-enhanced R-II Star Known}",
      journal = {\apjl},
         year = 2018,
        month = jun,
       volume = {859},
       number = {2},
          eid = {L24},
        pages = {L24},
          doi = {10.3847/2041-8213/aac722},
archivePrefix = {arXiv},
       eprint = {1805.11925},
 primaryClass = {astro-ph.SR},
       adsurl = {https://ui.adsabs.harvard.edu/abs/2018ApJ...859L..24H}
}

@ARTICLE{MP:Roederer2024,
       author = {{Roederer}, Ian U. and {Beers}, Timothy C. and {Hattori}, Kohei and {Placco}, Vinicius M. and {Hansen}, Terese T. and {Ezzeddine}, Rana and {Frebel}, Anna and {Holmbeck}, Erika M. and {Sakari}, Charli M.},
        title = "{The R-Process Alliance: 2MASS J22132050{\textendash}5137385, the Star with the Highest-known r-process Enhancement at [Eu/Fe] = +2.45}",
      journal = {\apj},
         year = 2024,
        month = aug,
       volume = {971},
       number = {2},
          eid = {158},
        pages = {158},
          doi = {10.3847/1538-4357/ad57bf},
archivePrefix = {arXiv},
       eprint = {2406.02691},
 primaryClass = {astro-ph.SR},
       adsurl = {https://ui.adsabs.harvard.edu/abs/2024ApJ...971..158R}
}

@ARTICLE{Chen2024MNRAS,
       author = {{Chen}, Meng-Hua and {Li}, Li-Xin and {Chen}, Qiu-Hong and {Hu}, Rui-Chong and {Liang}, En-Wei},
        title = "{Neutron star mergers as the dominant contributor to the production of heavy r-process elements}",
      journal = {\mnras},
         year = 2024,
        month = apr,
       volume = {529},
       number = {2},
        pages = {1154-1160},
          doi = {10.1093/mnras/stae475},
archivePrefix = {arXiv},
       eprint = {2402.08214},
 primaryClass = {astro-ph.HE},
       adsurl = {https://ui.adsabs.harvard.edu/abs/2024MNRAS.529.1154C}
}

@ARTICLE{Chen2025A&A,
       author = {{Chen}, Meng-Hua and {Li}, Li-Xin and {Liang}, En-Wei and {Wang}, Ning},
        title = "{Impact of nuclear mass models on r-process nucleosynthesis and heavy element abundances in r-process-enhanced metal-poor stars}",
      journal = {\aap},
         year = 2025,
        month = jan,
       volume = {693},
          eid = {A1},
        pages = {A1},
          doi = {10.1051/0004-6361/202451986},
archivePrefix = {arXiv},
       eprint = {2411.17076},
 primaryClass = {astro-ph.HE},
       adsurl = {https://ui.adsabs.harvard.edu/abs/2025A&A...693A...1C}
}

@ARTICLE{2007PhR...450...97A,
       author = {{Arnould}, M. and {Goriely}, S. and {Takahashi}, K.},
        title = "{The r-process of stellar nucleosynthesis: Astrophysics and nuclear physics achievements and mysteries}",
      journal = {\physrep},
         year = 2007,
        month = sep,
       volume = {450},
       number = {4-6},
        pages = {97-213},
          doi = {10.1016/j.physrep.2007.06.002},
archivePrefix = {arXiv},
       eprint = {0705.4512},
 primaryClass = {astro-ph},
       adsurl = {https://ui.adsabs.harvard.edu/abs/2007PhR...450...97A}
}

@ARTICLE{2022APPSB..32...19W,
       author = {{Wu}, Meng-Ru and {Banerjee}, Projjwal},
        title = "{The production of actinides in neutron star mergers}",
      journal = {Association of Asia Pacific Physical Societies Bulletin},
         year = 2022,
        month = dec,
       volume = {32},
       number = {1},
          eid = {19},
        pages = {19},
          doi = {10.1007/s43673-022-00048-1},
archivePrefix = {arXiv},
       eprint = {2205.11806},
 primaryClass = {astro-ph.HE},
       adsurl = {https://ui.adsabs.harvard.edu/abs/2022APPSB..32...19W}
}

@ARTICLE{Cescutti2014AAP,
  author = {{Cescutti}, G. and {Chiappini}, C.},
  title = {Explaining the {Ba}, {Y}, {Sr}, and {Eu} abundance scatter in metal-poor halo stars: constraints to the r-process},
  journal = {\aap},
  year = {2014},
  volume = {565},
  pages = {A51},
  doi = {10.1051/0004-6361/201423432}
}

@ARTICLE{Kobayashi2023ApJL,
  author = {{Kobayashi}, Chiaki and {Mandel}, Ilya and {Belczynski}, Krzysztof and {Goriely}, Stephane and {Janka}, Thomas H. and {Just}, Oliver and {Ruiter}, Ashley J. and {Vanbeveren}, Dany and {Kruckow}, Matthias U. and {Briel}, Max M. and {Eldridge}, Jan J. and {Stanway}, Elizabeth},
  title = {Can Neutron Star Mergers Alone Explain the r-process Enrichment of the {Milky Way}?},
  journal = {\apjl},
  year = {2023},
  volume = {943},
  pages = {L12},
  doi = {10.3847/2041-8213/acad82},
  archivePrefix = {arXiv},
  eprint = {2211.04964}
}

@ARTICLE{Lian2023MNRAS,
  author = {{Lian}, Jianhui and {Storm}, Nicholas and {Guiglion}, Guillaume and {Serenelli}, Aldo and {Cote}, Benoit and {Karakas}, Amanda I. and {Boardman}, Nick and {Bergemann}, Maria},
  title = {Observational constraints on the origin of the elements -- {VI}. Origin and evolution of neutron-capture elements as probed by the {Gaia-ESO} survey},
  journal = {\mnras},
  year = {2023},
  volume = {525},
  pages = {1329--1341},
  doi = {10.1093/mnras/stad2390},
  archivePrefix = {arXiv},
  eprint = {2308.01111}
}

@ARTICLE{2026SCPMA..6909513Q,
       author = {{Qiumu}, Wu-Zimo and {Chen}, Meng-Hua and {Chen}, Qiu-Hong and {Xie}, Fei and {L{\"u}}, Hou-Jun and {Wang}, Xiang-Gao and {Liang}, En-Wei},
        title = "{MeV gamma-ray lines from radioactive nuclei in magnetar giant flares}",
      journal = {Science China Physics, Mechanics, and Astronomy},
         year = 2026,
        month = aug,
       volume = {69},
       number = {10},
          eid = {109513},
        pages = {109513},
          doi = {10.1007/s11433-026-3045-7},
archivePrefix = {arXiv},
       eprint = {2606.17997},
 primaryClass = {astro-ph.HE},
       adsurl = {https://ui.adsabs.harvard.edu/abs/2026SCPMA..6909513Q}
}

	\bsp	
	\label{lastpage}
\end{document}